# Why high-error-rate random mutagenesis libraries are enriched in functional and improved proteins

D. Allan Drummond[1], Brent L. Iverson[2], George Georgiou[3],

and Frances H. Arnold[4]

[1]Program in Computation and Neural Systems, and

[4]Division of Chemistry and Chemical Engineering

California Institute of Technology, Mail Code 210-41

Pasadena, CA 91125-4100, USA

[2]Department of Chemistry and Biochemistry, and

[3]Department of Chemical Engineering

The University of Texas at Austin

1 University Station  A5300

Austin, TX  78712, USA

Corresponding author:   Frances H. Arnold
Division of Chemistry & Chemical Engineering
California Institute of Technology
Mail code 210-41
Pasadena, CA 91125
Tel:  (626) 395 4162
Fax: (626) 568 8743
E-mail:  frances@cheme.caltech.edu

# SUMMARY

Recently, several groups have used error-prone polymerase chain reactions to construct random mutagenesis libraries containing up to 27 nucleotide mutations per gene on average and reported a striking observation: although retention of protein function initially declines exponentially with mutations as has previously been observed, orders of magnitude more proteins remain functional at the highest mutation rates than this trend would predict. Mutant proteins having improved or novel activity were isolated disproportionately from these heavily mutated libraries, leading to the suggestion that distant regions of sequence space are enriched in useful positively coupled mutations and therefore that optimal mutagenesis should target these regions. If true, these claims have profound implications for laboratory evolution and for evolutionary theory. Here, we demonstrate that the distribution of mutations in high-error-rate error-prone PCR is not Poisson, as is often assumed, but a known distribution with broader variance. This has important consequences for measurements of protein mutational tolerance, of optimal mutation rates, and even of error rates themselves. We show that high-error-rate mutagenesis may be useful in certain cases for reasons unrelated to mutation coupling, and that optimal mutation rates are inherently protocol-dependent. Our results allow optimal mutation rates to be found given mutagenesis conditions and a protein of known mutational tolerance.

# INTRODUCTION

Laboratory evolution has been used to improve protein properties by mimicking natural evolution's stepwise exploration of sequence space[1], steadily improving protein activity or thermostability through repeated rounds of low-frequency mutation and selection. Because the fraction of proteins retaining function appears to decline exponentially with increasing numbers of amino acid substitutions[2,3,4,5], low mutation rates seek to create mutational diversity without destroying activity[6] so that improved clones can be found.

Recently, several groups reported construction of mutant libraries using high-mutation-rate error-prone polymerase chain reactions (EP-PCR) to probe distant regions of sequence space for an antibody fragment (up to an average $\langle m_{nt} \rangle$ = 22.5 nucleotide mutations per gene) [3,7], hen egg lysozyme (up to $\langle m_{nt} \rangle$ = 15.25) [8], and TEM-1 β-lactamase (up to $\langle m_{nt} \rangle$ = 27.2) [9]. Where both high and low error rates were assessed, the exponential trend in loss of function established for low $\langle m_{nt} \rangle$ was spectacularly violated at the highest rates, with orders of magnitude more functional clones isolated than would be expected[3,7,8]. Two studies reported improved or novel function more often in these high-mutation-rate libraries[3,9], leading to suggestions that low mutational pressure may not be optimal[3,9] and that hypermutagenesis can, without an exponentially increasing cost in inactivated sequences, explore multiple interacting mutations inaccessible to low-error-rate mutagenesis[9]. Note that these interactions could involve synergistic interactions to increase function directly, or they might involve combinations in which one or a few mutations increase function at the cost of folding or structural stability, the negative effects of which are suppressed by additional compensatory stabilizing mutations elsewhere in the protein.

The degree to which mutations interact, and thus mutational effects deviate from independence, is known as *epistasis*. Independent mutational effects imply an exponential decline in fraction functional with mutational distance, so the above studies' results suggest that mutations interact epistatically on average. Such a finding is of fundamental interest in evolutionary biology[10,11] and is potentially decisive in answering the major open question, "Why is there sex?"[12] Moreover, the discovery of reservoirs of positively interacting mutations would fundamentally change strategies for *in vitro* enzyme engineering by evolutionary methods[9]. Therefore, a careful analysis of these results is imperative.

Quantitative analysis of high-frequency mutagenesis results often assumes a Poisson distribution of mutations in error-prone PCR, an idea introduced by Shafikhani *et al.*[4]. This group's careful study on *B. lentus* subtilisin found an accurately reproducible exponential decline in fraction functional in all libraries where functional proteins were found, up to $\langle m_{nt} \rangle$=15, contrary to the upward trend reported later.

To examine the mutational distribution generated by high-error-rate error-prone PCR, we constructed two large libraries of single chain Fv (scFv) antibody mutants. The wildtype scFv antibody fragment derived from the 26-10 monoclonal antibody[13] binds digoxigenin with high affinity, and has been expressed as a fusion to the *E. coli* outer membrane protein Lpp-OmpA', allowing detection of mutants binding fluorescent-dye-conjugated digoxigenin by fluorescence-activated cell sorting (FACS)[3]. Libraries were assayed for mutant retention of wildtype affinity for digoxigenin (briefly, retention of function). These libraries were constructed and assayed exactly as in a previous study by two of the present authors[3], making the results of both studies directly comparable. We were able to determine how the mutational statistics relate to PCR experimental parameters and to retention of function.

We show that error-prone PCR at high error rates does not produce Poisson-distributed mutations, but follow a previously proposed distribution derived from a model of the actual PCR process[14]. We derive the expected fraction of functional mutants based on this more realistic model and show that many reported experimental mutation data follow this model's predictions. We then introduce a simple measure of optimality to evaluate optimal mutation rates for improvement of protein function. Our results suggest that the trends observed in earlier studies do not constitute evidence for positive epistasis.

## RESULTS

**Distribution of mutations generated by error-prone PCR**

The probability $\Pr(f)$ that an error-prone PCR-amplified sequence retains function can be obtained as follows. Sun[14] modeled error-prone PCR by assuming $n$ thermal cycles during which DNA strands are duplicated with probability $\lambda$, the PCR efficiency (assumed constant, realistic for large amounts of starting template[15,16]), resulting in $d=n\lambda$ DNA doublings and an average of $\langle m_{nt} \rangle$ nucleotide mutations per sequence. The mutational distribution under these assumptions can be written[14], with $x = \dfrac{\langle m_{nt} \rangle (1+\lambda)}{n\lambda}$, as

$$\Pr(m_{nt}) = (1+\lambda)^{-n} \sum_{k=0}^{n} \binom{n}{k} \lambda^k \frac{(kx)^{m_{nt}} e^{-kx}}{m_{nt}!}, \qquad (1)$$

which has mean $\langle m_{nt} \rangle$ and variance $\sigma^2_{m_{nt}} = \langle m_{nt} \rangle + \frac{\langle m_{nt} \rangle^2}{n\lambda} = \langle m_{nt} \rangle \left(1 + \frac{\langle m_{nt} \rangle}{d}\right)$. At large $\langle m_{nt} \rangle$, deviation from the Poisson assumption that the variance is equal to the mean $\langle m_{nt} \rangle$ can be profound. We call Equation 1 the *PCR distribution*.

**Results of mutagenesis**

To examine the mutational distribution generated by high-error-rate error-prone PCR, for which the Poisson- and PCR-based models make distinct predictions, we sequenced 45+ clones from each library. Though the same conditions were used to generate libraries A and B, different fractions of functional clones resulted.

Poisson-distributed mutations will have equal mean and variance, while PCR-distributed mutations will always have a variance larger than the mean. Figure 1 shows the distribution of nucleotide mutations observed in library A (46 sequences) and library B (45 sequences); summary statistics are shown in Table 1, and mutational spectra are reported in Table 2.

While visual inspection of the mutation histograms overlayed with the theoretical distributions cannot distinguish between the two models, the relevant statistics are stark and favor the PCR distribution while rejecting the Poisson distribution. For library A, $\langle m_{nt} \rangle$ = 15.8 and $\sigma^2_{m_{nt}}$ = 26.3; for library B, $\langle m_{nt} \rangle$ = 19.8 and $\sigma^2_{m_{nt}}$ = 36.1 (Table 1). The probability of measuring variances at least this large given an underlying Poisson distribution with the observed mean is < 0.005 for library A and < 0.001 for library B; the joint probability of observing two libraries with variances this high is < $10^{-5}$. With a PCR efficiency of $\lambda$ = 0.6 (18 doublings), the PCR distribution yields expected variances of 29.6 (library A) and 41.4 (library B), consistent with the observed values.

Using a likelihood ratio test on the mutational histograms, we reject the Poisson distribution in favor of the PCR distribution with two additional degrees of freedom ($n$ and $\lambda$) for library A ($\chi^2 = 7.39$, $P < 0.025$) and for library B ($\chi^2 = 8.63$, $P < 0.025$). (Using two additional degrees of freedom is conservative, since $n$ is fixed in each experiment.) Thus, the modeled PCR distribution better describes the data than the previously assumed Poisson model.

**Retention of protein function after mutation**

What is the effect of the non-Poisson mutational distribution on the fraction of clones in a library that retain function? We assume the probability an individual protein will retain function after $m_{aa}$ amino acid substitutions declines exponentially according to $\Pr(f|m_{aa}) = v^{m_{aa}}$, where $v$ can be interpreted as the average fraction of functional one-mutant neighbors on the sequence-space network[10,17]; this assumption is consistent with experimental results obtained without using PCR[2] and with theoretical considerations[5]. Note that this model assumes no average epistasis. The probability a nucleotide mutation produces an amino acid change is assumed to be binomial with parameter $p_{ns}$, the probability of a nonsynonymous mutation, corresponding to the assumption that mutations hit distinct codons. This assumption and the value $p_{ns} = 0.7$ appear realistic[3]. Insertions, deletions, and mutations to stop codons truncate and inactivate the encoded protein; these we assume occur at a rate $p_{tr}$ on the order of 3-5%[4]. The probability a sequence with $m_{nt}$ nucleotide mutations retains function includes all these effects and is then

$$\Pr(f|m_{nt}) = (1-p_{tr})^{m_{nt}} \sum_{m_{aa}=0}^{m_{nt}} \binom{m_{nt}}{m_{aa}} p_{ns}^{m_{aa}} (1-p_{ns})^{m_{nt}-m_{aa}} v^{m_{aa}} \qquad (2)$$
$$= ((1-p_{tr})(1-(1-v)p_{ns}))^{m_{nt}}.$$

Under the assumption of Poisson-distributed mutations, Shafikhani *et al.*[4] showed that, if a fraction $q_i$ of nucleotide mutations inactivate a protein, the fraction functional declines exponentially as $e^{-\langle m_{nt}\rangle q_i}$. Because $q_i = p_{tr} + (1-p_{tr})(1-v)p_{ns}$, we expect $\Pr(f) = e^{-\langle m_{nt}\rangle(p_{tr}+(1-p_{tr})(1-v)p_{ns})}$ in a Poisson-distributed library. This exponential decline became the experimental expectation for subsequent groups, leading to surprise when functional mutants were later found in great excess at high average mutation rates. By combining Equations (1) and (2) and assuming gene length $L \to \infty$, a mild assumption when $\langle m_{nt}\rangle \ll L$, we find the probability a sequence from the library will retain function is

$$\Pr(f) = \sum_{m_{nt}=0}^{\infty} \Pr(f|m_{nt})\Pr(m_{nt}) = \left(\frac{1 + \lambda e^{-\frac{\langle m_{nt}\rangle(1+\lambda)}{n\lambda}(p_{tr}+(1-p_{tr})(1-v)p_{ns})}}{1+\lambda}\right)^n. \quad (3)$$

Equation 3 makes several predictions. In the limit of many thermal cycles $n$, all else equal, the original expectation $\Pr(f) = e^{-\langle m_{nt}\rangle(p_{tr}+(1-p_{tr})(1-v)p_{ns})}$ (above) is recovered. If the number of thermal cycles $n \propto \langle m_{nt}\rangle$, following the protocol of Shafikhani *et al.*, then $\Pr(f)$ should be a perfect exponential in $\langle m_{nt}\rangle$, which is precisely what this group reports. However, if $n$ is fixed as in other studies[3,8,9], then $\Pr(f)$ curves upward relative to an exponential decline as $\langle m_{nt}\rangle$ increases. PCR efficiency $\lambda$ decreases with increasing $\langle m_{nt}\rangle$[18], which increases the expected curvature. In other words, there will be more functional sequences than predicted by the exponential decline.

Using the previously reported scFv antibody data[3] for low $\langle m_{nt}\rangle$, where the Poisson assumption is not unreasonable, and the reported value $q_i = 0.6 \approx (1-v)0.7$, we can estimate $v \approx 0.2$ for the antibody binding task. For the subtilisin data[4], we similarly use the reported $q_i = 0.27 \approx (1-v)0.7$ to estimate $v \approx 0.6$. With these values for $v$, Figure 2 compares the predictions of Equation 3 to the observed fractions of functional clones at various library mutation levels $\langle m_{nt} \rangle$ reported by Daugherty *et al.*[3] and in the present work for the scFv antibody fragment (Fig. 2a) (see also Table 3) and Shafikhani *et al.*[4] for subtilisin (Fig. 2b). The agreement is quite good and demonstrates that the excess of functional clones can in fact be consistent with an underlying exponential relationship between number of amino acid substitutions and probability of retained wild-type function. To further test our analytical predictions, we simulated single-round error-prone PCR using template DNA strands encoding a folded "wildtype" lattice protein. The amplified DNA was translated into lattice proteins which were scored as functional if they retained the fold and thermostability of the wildtype. We observed excellent agreement with Equation 3 (see Supporting Information).

The reason for deviation from an exponential decline is hinted at in the limit of large average mutation rates, when the exponential part of Equation 3 vanishes and $\Pr(f)$ approaches a constant, $\Pr(f) \to (1+\lambda)^{-n}$. For a mutationally fragile protein such as the scFv antibody performing the digoxigenin binding task, this can occur at experimentally accessible mutation rates, as can be seen most clearly in the library originally reported[3] and revisited by Georgiou[7]. As the mutation rate increases, the antibody fragment becomes "quite insensitive to mutational load" and $\Pr(f)$ flattens out at a value of roughly 0.0018 [7]. Most interestingly, this limiting value is a function only of the PCR conditions, and does not depend on the protein at all.

What causes these counterintuitive results? Error-prone PCR at high frequency generates heavily mutated sequences by a process akin to Xeroxing copies of copies: low-fidelity copies give rise to even lower-fidelity copies, yet a copy, once produced, is not replaced, but remains in the final distribution of copies. During the polymerase chain reaction, the first generation of mutants, amplified directly from the wild-type template gene and carrying few mutations, persists in the mix and continues to reproduce copies with few additional mutations throughout subsequent cycles. The protein products of these less-mutated copies retain function at a greatly elevated rate compared to the average sequence, leading to upward bias in the functional fraction.

**Why are improved mutants found more often in high-error-rate libraries?**

If statistical effects of the mutagenesis protocol can explain the dramatic deviation from exponential in the fraction of functional sequences without recourse to epistasis, why are high-$\langle m_{nt} \rangle$ libraries enriched in improved clones, despite a smaller number of clones retaining *any* function? To address this question, we now explore another consequence of PCR's broad mutational distribution.

The effective size of a library is not the number of mutants screened, the number usually reported, but rather the number of *unique* mutants screened. In a library of $10^6$ transformants of the scFv antibody gene (726 bp, 242 aa) with an average of one mutation per sequence, most of the 2,178 possible 1-mutants will occur on the order of 100 times, reducing the effective library size by roughly two orders of magnitude. Most mutagenesis is concerned with protein sequences, where additional losses occur. Truncations due to frameshift mutations or mutations to stop codons eliminate a significant fraction of sequences. With one nucleotide mutation per codon, an average of 5.7 amino acid substitutions (out of a maximum of 19) are accessible due to the conservatism of the genetic code, for a total of $242 \times 5.7 = 1,379$ accessible amino acid sequences with one substitution. (We ignore the effects of synonymous mutations.) One million transformants thus yield just over one thousand unique protein sequences, about a 1,000-fold reduction in the effective library size. The problem of mutant recurrence raises the question: What is the number of unique protein sequences in an error-prone PCR library?

We estimate the number of unique sequences in the following way. We derive the distribution of numbers of amino acid substitutions $\Pr(m_{aa})$ after error-prone PCR, estimate the number of non-truncated sequences $N_{m_{aa}}$ with each $m_{aa}$ in a library of a given size, compute the expected number of unique sequences $U_{m_{aa}}$ at each $m_{aa}$ by accounting for recurrence among the $N_{m_{aa}}$ sequences, and then find the expected number of unique sequences $U$ by simply summing the $U_{m_{aa}}$.

Given PCR conditions as before with an average number of nucleotide mutations per sequence $\langle m_{nt} \rangle$, what is the distribution of the number of amino acid substitutions per sequence $\Pr(m_{aa})$? We assume, as before, that each nucleotide mutation changes the encoded amino acid with probability $p_{ns}$, so we obtain

$$\Pr(m_{aa}) = \sum_{m_{nt}=m_{aa}}^{L} \Pr(m_{nt}) \binom{m_{nt}}{m_{aa}} p_{ns}^{m_{aa}} (1-p_{ns})^{m_{nt}-m_{aa}}$$
$$= (1+\lambda)^{-n} \sum_{k=0}^{n} \binom{n}{k} \lambda^k \frac{(ky)^{m_{aa}} e^{-ky}}{m_{aa}!} \quad (4)$$

with $y = \dfrac{\langle m_{nt} \rangle p_{ns}(1+\lambda)}{n\lambda}$. That is, the distribution of amino acid substitutions $\Pr(m_{aa})$ is equivalent, in form, to the distribution of nucleotide mutations $\Pr(m_{nt})$, but with an average of $\langle m_{aa} \rangle = \langle m_{nt} \rangle p_{ns}$ substitutions. For simplicity, we will drop the subscript for amino acid substitutions, and use $m$.

Of the sequences with $m$ amino acid substitutions, some will also be truncated by frameshifts or stop codons. Treating all truncations as nonsynonymous changes, the fraction of non-truncated sequences with $m$ substutions is $\Pr(\text{nontruncated}|m) = (1 - p_{tr}/p_{ns})^m$. Given an error-prone PCR library of $N$ transformants,

$$N_m = N \Pr(m) \Pr(\text{nontruncated}|m) \tag{5}$$

on average are non-truncated proteins with $m$ amino acid substitutions.

Of these proteins with $m$ substitutions, how many unique proteins could we possibly find? If $m = 0$, at most one unique sequence exists, even if thousands are generated. For any $m$ there are on average $M_m = \binom{L/3}{m} 5.7^m$ total unique proteins accessible assuming at most one mutation per codon, where $L$ is the length of the gene in nucleotides.

Given $N_m$ samples, how many of these $M_m$ unique proteins can we expect to find? This is the classic "coupon collector problem"[19] and directly addresses the question of mutant recurrence, since any sample either yields a new, unique protein or one that has been sampled before. The expected number of unique sequences produced by equiprobably sampling $M_m$ sequences $N_m$ times is

$$U_m = M_m - M_m(1 - 1/M_m)^{N_m} \approx M_m(1 - e^{-N_m/M_m}). \tag{6}$$

For example, to sample all $M_m = 1{,}379$ accessible 1-mutants of scFv requires 8-fold oversampling, $N_m = 11{,}000$ samples. Taking only 1,379 samples, $N_m = M_m$, yields 872 unique proteins, 63% of the total.

In practice, for proteins of a few hundred amino acids and libraries of a few million transformants, recurrence need only be considered for small values of $m$ ($m<3$), because sequence space becomes large enough to make recurrence extremely unlikely at higher $m$ values so that $U_m \approx N_m$.

The total number of unique sequences in a library is simply the sum over all unique sequences with a specific number of substitutions:

$$U = \sum_{m=0}^{L/3} U_m. \tag{7}$$

Figure 3a shows the fraction of unique sequences $U/N$ obtained from simulations (see Methods) in which the scFv gene was mutated according to PCR statistics with the observed frequencies (Table 2, with 3% frameshift rate) or unbiased frequencies (all mutations equally weighted, with 3% frameshift rate). The prediction from Equation 7 is also plotted and agrees well. Increasing the mutation rate increases the number of unique sequences because fewer are lost to recurrence. Note that even at the highest mutation rates, the fraction of unique sequences does not approach 1.0, because sequences truncated by frameshifts and stop codons are not considered unique and accumulate at increasing levels as the mutation rate is increased.

Of greater interest is the expected number of unique sequences in the library that are expected to retain at least wildtype function, because these sequences are a superset of potentially improved sequences. We can estimate the number of unique, functional sequences as

$$U_f = \sum_{m=0}^{L/3} U_m v^m. \tag{8}$$

Figure 3b shows the fraction of unique, functional sequences $U_f/N$ obtained from the same simulations as in Fig. 3a, with Eq. 8 plotted for comparison. Biases in mutation frequencies decrease the fraction of unique sequences but preserve the overall form. Results using unbiased frequencies are predicted accurately by our theoretical treatment.

Clearly, low-error-rate libraries suffer from dramatic mutant recurrence, an effect avoided at high error rates. Improved proteins are found often in high-error-rate libraries *because these libraries contain more unique functional sequences*.

**Optimal random mutagenesis**

A typical and important goal in protein engineering is to improve an existing protein function, for example by increasing catalytic rate, thermostability, binding affinity, or specificity. While rational engineering has made significant strides, high-throughput screening of large mutant libraries for improved clones is both a dominant strategy to achieve this goal and an area of active research[7].

Given a choice of protein scaffold, a library of fixed size, and no reliable basis for rational engineering, a simple measure of library optimality is the number of unique functional sequences. Figure 3b shows that, given this measure, an optimal mutation rate exists which balances diversity (uniqueness is lost if $\langle m_{nt} \rangle$ is too low) with retained function (functional sequences are rare if $\langle m_{nt} \rangle$ is too high). Mutational biases do not significantly affect the optimal mutation rate.

The optimum depends on the number of transformants screened and on the PCR protocol used, among other parameters. Figure 4a compares predicted optimal mutation rates under identical PCR conditions for the scFv antibody ($v \approx 0.2$) depending on whether a thousand or a million clones are screened. The difference, 1.3 average nucleotide substitutions, corresponds to one amino acid substitution on average. Figure 4b compares predicted optimal mutation rates under identical conditions and with the same wildtype protein, but using 30 cycles (as in the present work) in one case and 2 cycles (as in Ref. 9) in the other. A difference of one nucleotide mutation results. Optimal rates also depend on protein mutational tolerance as reflected by $v$: the more tolerant the protein, the higher the optimal mutation rate (not shown).

Table 3 lists estimates for $U_f$ given the experimental conditions reported here and previously[3]. Despite the over 200-fold lower observed percentage of functional transformants isolated from the highest-$\langle m_{nt} \rangle$ library relative to the lowest, and the 14-fold fewer functional sequences observed, only 60% fewer unique functional sequences are expected in the highest-$\langle m_{nt} \rangle$ library. Given the experimental parameters of the highest-$\langle m_{nt} \rangle$ library and altering only the mutation rate, the rate $\langle m_{nt} \rangle$ = 11.5 is predicted to produce more unique functional sequences (>10,000) than any of the reported libraries. The *optimal* mutation rate given the highest-$\langle m_{nt} \rangle$ experimental parameters is predicted to be around $\langle m_{nt} \rangle$ = 3.1, which is predicted to yield >35,000 unique, functional sequences.

## DISCUSSION

Laboratory evolution by random mutagenesis remains the most effective known strategy for improving enzyme properties given a choice of scaffold and no reliable basis for rational engineering. The possibility that distant regions of sequence space harbor excesses of functional and, for at least some enzymatic tasks, improved proteins has been advanced several times, with significant experimental evidence to bolster the claims. We have shown that a more accurate model of error-prone PCR than previously used, due to Sun[14], is required to adequately describe the mutational distribution resulting from high-error-rate error-prone PCR. This model, in turn, provides straightforward explanations for the previously observed experimental findings: 1) the excess functional proteins observed at high $\langle m_{nt} \rangle$ is predictable using our Equation 3, is due to low-mutation sequences generated early in the reaction, and is consistent with an exponential decrease in retention of function with amino acid substitution level; and 2) loss of functional sequences at high mutation rates can be balanced by diversity in the form of more unique sequences, improving sampling of sequence space and leading to a higher probability that improved mutants will be found if they exist. We have demonstrated the often-overlooked importance of accounting for recurrence of mutants when estimating how much of sequence space a library covers, extending previous work on modeling effects of mutational bias[20]. Using our simple definition of library optimality as maximizing the number of unique, functional proteins, these two observations lead to an optimal mutation rate which can be estimated using our analytical results. However, optimal mutation rates are both protocol- and protein-dependent. Optimal rates derived for error-prone PCR using one set of conditions do not necessarily hold for another set (Fig. 4), and are highly unlikely to hold for saturation mutagenesis or site-directed mutagenesis, for which uniqueness is rarely a problem and the distribution of mutation levels in a typical library is tight and easily controllable.

We have explained several disparate mutagenesis results using only a single parameter unrelated to experimental protocols: $v$, the average probability of retaining wildtype function after a random amino acid substitution[5]. It follows that these experiments can be used to measure $v$ using the analytical tools we have introduced here, with an important caveat. Because multiple mutations per codon, rarely found in error-prone PCR even at high mutation rates (though not always[21]), are necessary to experimentally measure $v$, such experiments cannot directly measure this parameter but can provide a credible upper bound due to the conservative nature of the genetic code. While $v$ relates simply to the "structural plasticity" $q_i = p_{tr} + (1-p_{tr})(1-v)p_{ns}$ proposed by Shafikhani *et al.* [4], our results show the emergence of a perfect exponential decline in their experiments likely depended both on a fundamental property of proteins and the particular experimental protocol employed. We also distinguish between genetic mutations which produce truncated protein products, essentially all of which lack function, and those which produce full-length proteins whose structural properties determine whether mutations are tolerated. We believe $v$ more accurately captures the idea of *structural* plasticity.

A protein's intrinsic functional tolerance to substitutions is only one of many ways in which genetic mutations may affect the fraction of active clones in a library. Biologically relevant or screenable activity may depend on the action of many molecules in an organism, so mutations which hinder expression (e.g. through introduction of non-preferred codons, or in rarer cases by altering mRNA secondary structure) may decrease the fraction of clones scored as active. Disruption of signal sequences may result in improper targeting to cellular locations such as the periplasm or cell membrane. Mutations may destabilize the protein, hindering its folding or exposing it to proteolysis or irreversible misfolding without actually destroying the function of the natively folded molecule. The dominant effect of most random mutagenesis is changes in the primary sequence of a target protein, most of which disrupt native function, and our simple treatment appears to work well under these circumstances.

Our results also illuminate potentially serious methodological flaws in previous studies. For example, the accuracy in measuring average library mutation rate by nucleotide sequencing depends on the variance of the mutational distribution, which at high mutation rates is far broader than that of the Poisson distribution previously assumed. The expected standard error of measurement on a library with $\langle m_{nt} \rangle$ average mutations assessed by sequencing $N_{seq}$ clones is $\sigma_m / \sqrt{N_{seq}} = \sqrt{\langle m_{nt} \rangle (1 + \langle m_{nt} \rangle / n\lambda) / N_{seq}}$. Zaccolo and Gherardi[9], for example, report four libraries averaging $\langle m_{nt} \rangle$ = 8.2, 19.7, 21.3 and 27.2 mutations per coding region of a 1,088 base-pair gene constructed using 2, 5, 10 and 20 thermal cycles with $\langle m_{nt} \rangle$ measured by sequencing at least 2,500 base pairs, effectively $N_{seq}$ = 2.5. Even if the true value of $\langle m_{nt} \rangle$ is as measured and perfect PCR efficiency assumed, these measurements have an expected $1\sigma$ standard error of 4.3, 6.5, 5.4 and 5.3 mutations per gene, respectively, calling into question the actual levels of hypermutagenesis achieved in these experiments.

The analysis presented here has important consequences for understanding the natural and directed evolution of proteins. Importantly, we have provided a thorough analysis of an apparent manifestation of mutational epistasis.

Two issues are often confused: whether mutations interact epistatically on average in *individual* folded sequences, and whether mutations interact epistatically on average in a library or *ensemble* that contains both folded and unfolded sequences. Ensemble epistasis is the only measure of interest in studies of the evolutionary persistence of sexual recombination[12] and of primary interest in deciding which regions of sequence space should be targeted for efficient directed evolution.

If ensemble epistasis existed, as implied by earlier interpretations of the less-than-exponential decline in retention of function with mutational distance discussed in the present work, then individual epistasis would also be found on average. Importantly, the reverse is not true. Though folded or improved proteins may display cooperative effects (mutations which are better together than individually), many polypeptides in a random library may also carry mutations that are more deleterious together than apart.  However, the latter are unlikely to be found by investigators, because such mutants are disproportionately likely to fail to fold, and little if any attention is given to the vast numbers of unfolded proteins in mutant libraries.  Confusion arising from the asymmetry between types of epistasis—ensemble epistasis implies individual epistasis, but individual epistasis does not imply ensemble epistasis—may have inspired prior claims that high mutation rates can be used to access reservoirs of cooperative mutations while only a "small proportion" of clones will be lost to disruptive mutations[9].

As a result of our analysis, several data sets probing high mutation rates can now be seen, despite appearances to the contrary, to provide no evidence for ensemble epistasis, of particular biological interest given the recent discoveries of multiple native error-prone polymerases in bacteria and higher organisms[22]. Meanwhile, recent work providing a explanation for why the fraction of mutant proteins retaining function will decline exponentially[5] suggests that ensemble epistasis is unlikely.  We cannot rule out the existence of epistasis; our analysis merely points out one way in which a mutation process can produce results which give the *appearance* of epistasis when there is none.

Exploration of distant regions of sequence space by random mutation alone appears highly inefficient, reinforcing the role of other search process such as homologous recombination in creating sequence diversity[23,24].  High-mutation-rate error-prone PCR, however, can be used to overcome the "uniqueness sink" that occurs at low mutation rates when using selection or high-throughput screening to assay large numbers of clones.  Finally, optimal mutation rates cannot be decoupled from the physical process of mutation, making them dependent on the particular organism or protocol under consideration. There can be no "optimal mutational load for protein engineering," as has previously been suggested[21], without specification of the engineering methodology.

## MATERIALS AND METHODS

**Library construction, sequencing and functional assay**

We constructed two libraries, A and B, from error prone PCR reactions as described.[18] Identical mutagenesis conditions were used for both libraries but produced different mutation levels in each library. In particular, 2.50 mM MgCl$_2$, 0.5 mM MnCl$_2$, 0.35 mM dATP, 0.40 mM dCTP, 0.20 dGTP, and 1.35 mM dCTP were used along with Taq DNA Polymerase. The PCR reaction was continued for 30 cycles rather than 16 as in the reference. All other parameters, and subsequent ligation, transformation and FACS functional analysis procedures were performed as previously described.[3]

**Simulation**

To simulate the error-prone PCR process, two approaches were taken. First, we exhaustively simulated the error-prone PCR process using genes encoding simplified model proteins (compact lattice model, 25 residues consisting of any of 20 amino acids) which were then folded and assayed for retention of wildtype structure. Details of this simulation and results are presented in Supporting Information.

We found that a vastly simpler simulation, produced nearly identical results (see Supplemental Figure S2) and used this simulation to generate Fig. 3. The scFv gene was mutated $N = 50{,}000$ times at each $\langle m_{\rm nt} \rangle$ according to the observed mutation frequencies (Table 2, Library A) and the PCR distribution, Equation 1, with parameters as indicated in the figure legend. Each mutated gene was translated into a protein sequence according to the universal genetic code. Truncated proteins, either from stop codons or frameshifts, were discarded. Whether a full-length sequence was functional or not was estimated by counting the number of amino acid substitutions relative to wildtype and designating the protein functional with probability $\Pr(f|m_{aa}) = v^{m_{aa}}$. All full-length protein sequences were inserted in a set which retained only unique sequences. Numbers and fractions of unique, functional and jointly unique and functional sequences were then tabulated.

## ACKNOWLEDGMENTS

We thank G. Chen and R. Loo for creation and screening of the scFv libraries, J. D. Bloom for optimal mutation rate calculations, C. C. Adami for guidance, C. O. Wilke for lattice protein folding code and advice, and Z.-G. Wang for insightful comments on the manuscript. D.A.D. acknowledges NIH National Research Service Award 5 T32 MH19138. This research is supported by Army Research Office Contract DAAD19-03-D-0004.

# FIGURE LEGENDS

**Figure 1.** Mutational distributions for two high-error-rate scFv antibody libraries compared with Poisson and PCR distributions. **a**: Library A, 46 sequences. **b**: Library B, 45 sequences. The corresponding PCR distributions with the same means (see Table 1) (solid line, $n = 30$ cycles and efficiency $\lambda = 0.6$) and Poisson distribution (dashed line) are shown for comparison. For these histograms, the Poisson distribution may be rejected in favor of the PCR distribution (see text).

**Figure 2**. Equation 3 explains previously reported experimental results. **a**: Comparison to scFv antibody data from Daugherty et al.[3] (□) and present work (■); for conditions, see the footnotes to Table 3. Dashed line is the original fit reported[3], $e^{-\langle m_{nt}\rangle q_i}$ with $q_i = 0.6$. Solid lines show Eq. 3 for the two libraries reported here (bottom) and for the highest-$\langle m_{nt}\rangle$ library conditions reported previously[3] (top). Changes in line curvature are due entirely to changes in PCR efficiency $\lambda$. **b**: Comparison to high-$\langle m_{nt}\rangle$ subtilisin data from Shafikhani et al.[4] (open squares with standard error bars), which were produced by a multi-round protocol. Conditions (all per-round): $d = n\lambda = 10$ DNA doublings, $n=13$ thermal cycles, $\langle m_{nt}\rangle = 2.01$ or 5.17 nucleotide mutations per gene. The fractions functional predicted by Eq. 3 for a multi-round protocol (solid line) and a single-round protocol (dotted line) show that the theory properly predicts the observed exponential decline in fraction functional.

**Figure 3**. Error-prone PCR error rates strongly influence the fraction of unique and functional sequences. **a**: Fraction of unique sequences in a simulated library

of 50,000 scFv clones ($v = 0.2$) using the observed mutational spectrum ($\diamond$) or an unbiased spectrum ($\blacklozenge$). Line is Eq. 7 evaluated with $n = 30$ thermal cycles, efficiency $\lambda = 0.6$, $p_{ns} = 0.76$ and $p_{tr} = 0.07$. **b**: Fraction of unique and functional sequences in the same library. Line is Eq. 8 evaluated using the same parameters. An optimal mutation rate exists which balances uniqueness with retention of function. Mutational biases lower the fraction of unique and functional sequences, but do not significantly alter the optimal mutation rate.

**Figure 4**. The requirement for uniqueness reduces effective library size and leads to library- and protocol-dependent optimal library mutation rates. **a**: Optimal mutation rate ($\bullet$) depends on library size. Predicted fractions of unique functional sequences given by Eq. 8 for the same protocol ($n = 30$ thermal cycles with efficiency $\lambda = 0.6$, $p_{ns} = 0.76$ and $p_{tr} = 0.07$) and protein (scFv-like, $v = 0.2$) are shown at each average mutation rate $\langle m_{nt} \rangle$ if $10^3$ transformants (top, $\langle m_{nt} \rangle_{opt} = 1.5$) or $10^6$ transformants (bottom, $\langle m_{nt} \rangle_{opt} = 2.8$) are screened. **b**: Optimal mutation rate ($\bullet$) depends on PCR protocol. Predicted fractions of unique functional sequences given by Eq. 8 are shown for the same protein (scFv-like, $v = 0.2$) and library size ($10^5$ transformants) using $n = 30$ thermal cycles (top, $\langle m_{nt} \rangle_{opt} = 2.8$) or $n = 2$ thermal cycles (bottom, $\langle m_{nt} \rangle_{opt} = 1.8$). In all cases, recurrence leads to profound loss of uniqueness at low $\langle m_{nt} \rangle$, and optimal $\langle m_{nt} \rangle$ balances uniqueness and retention of function.

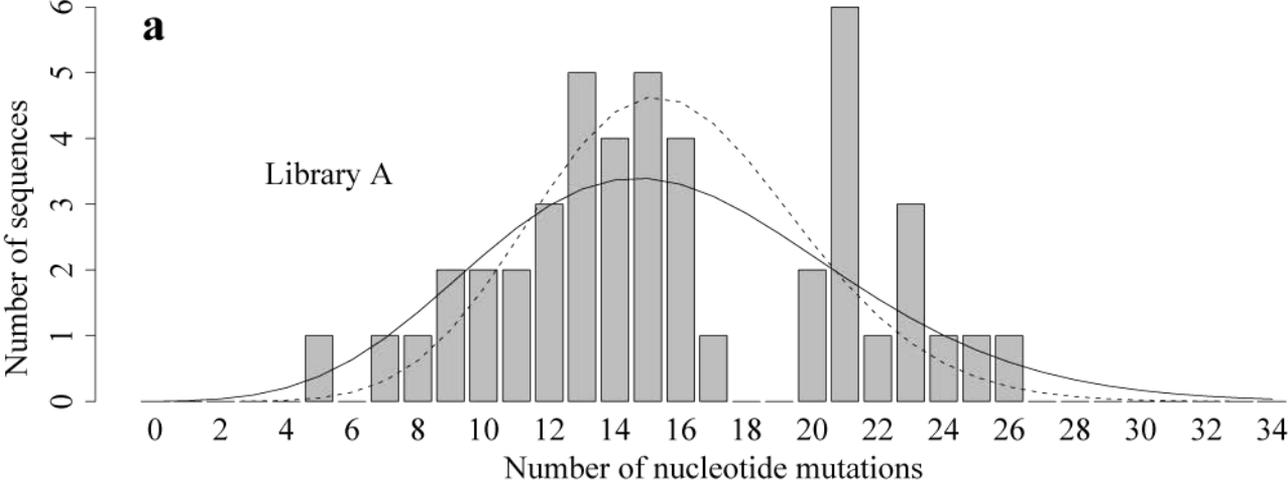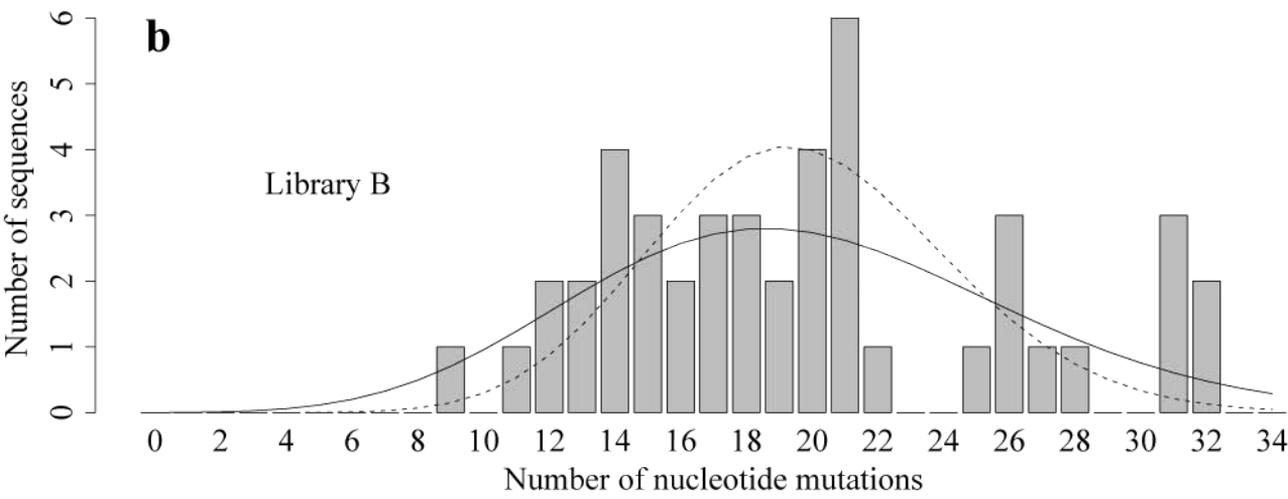

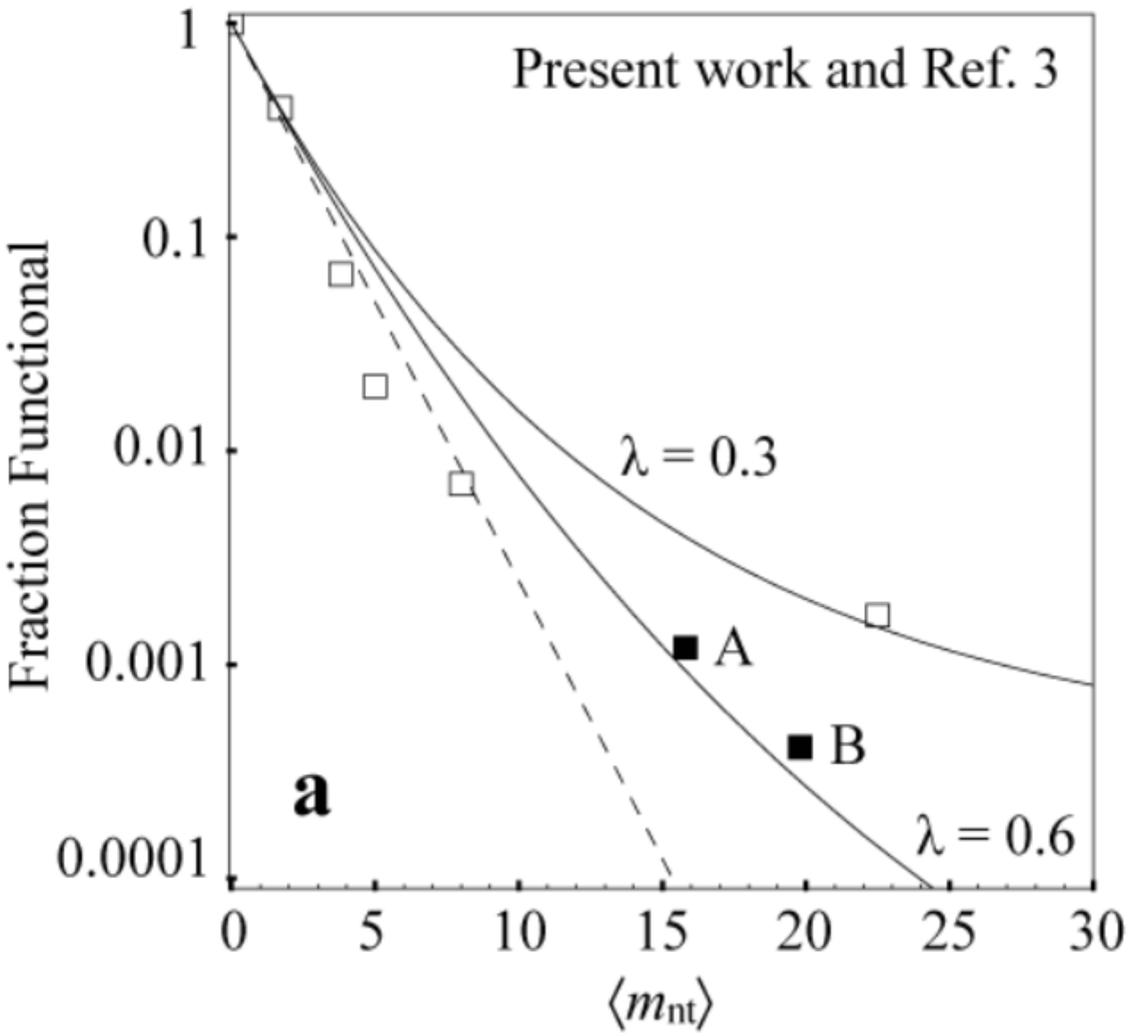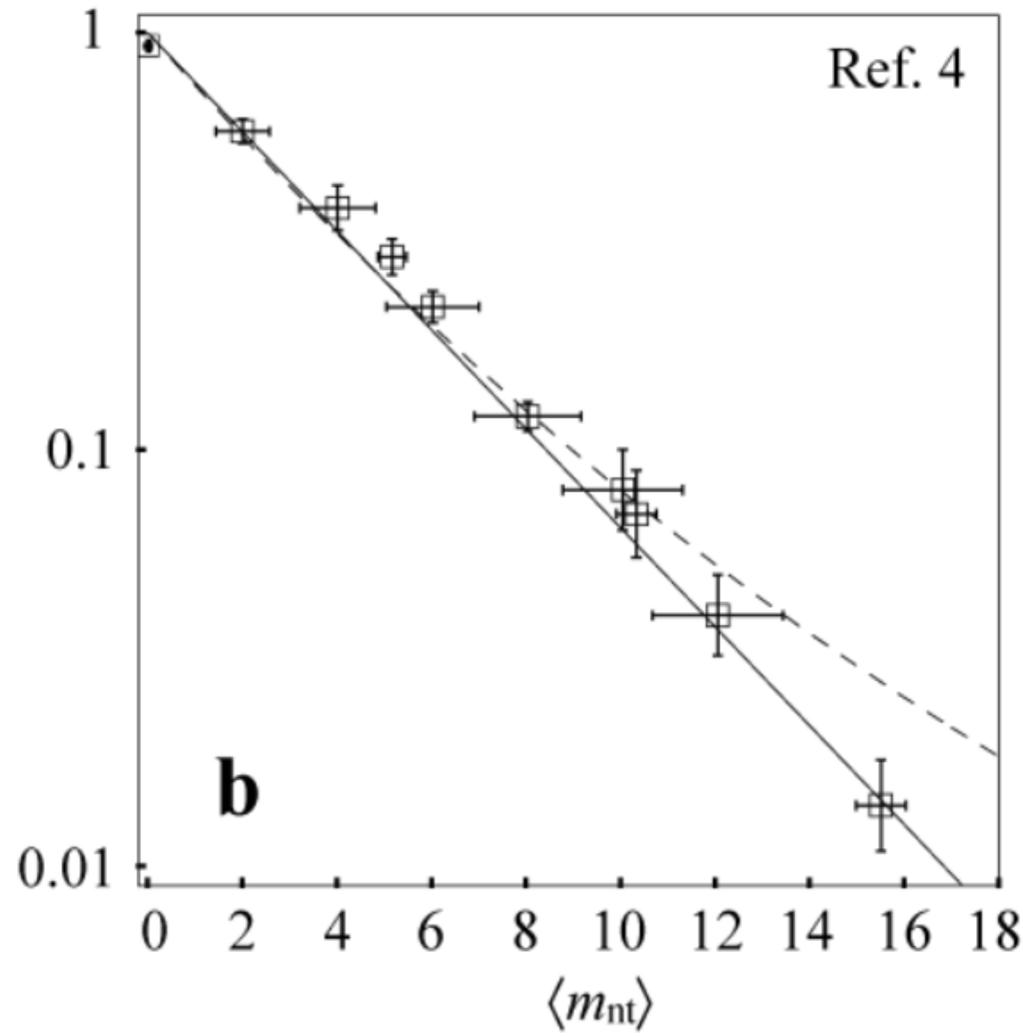

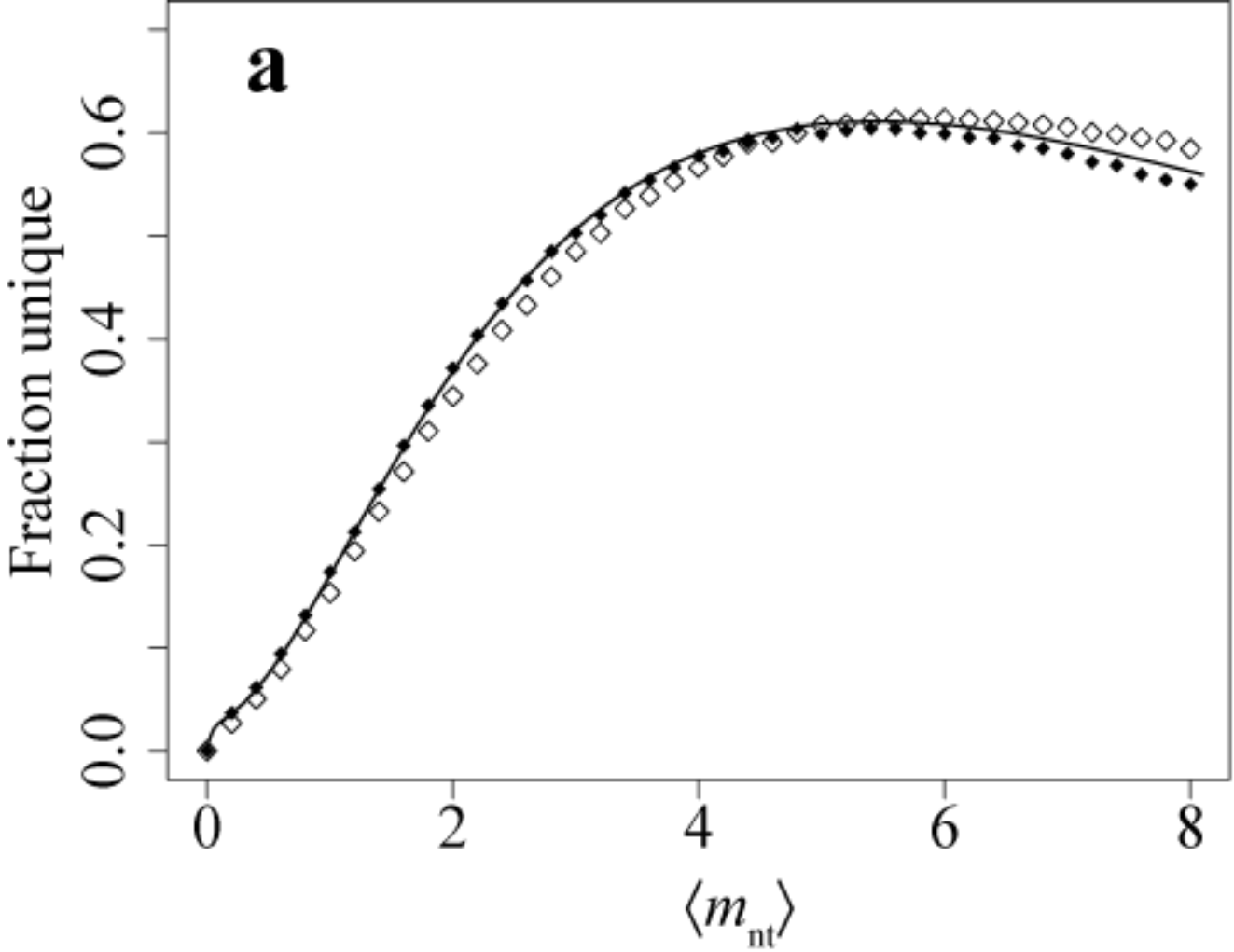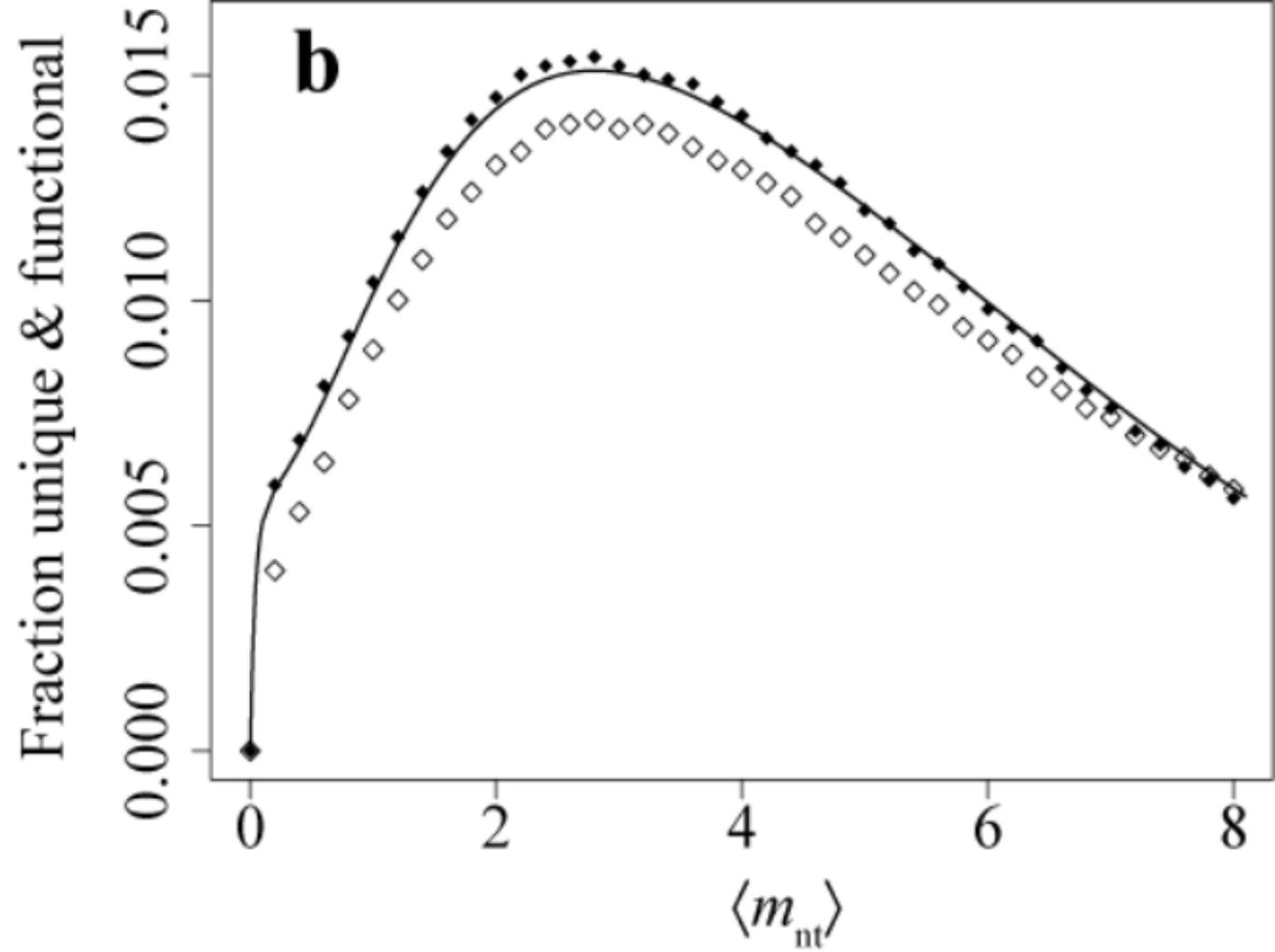

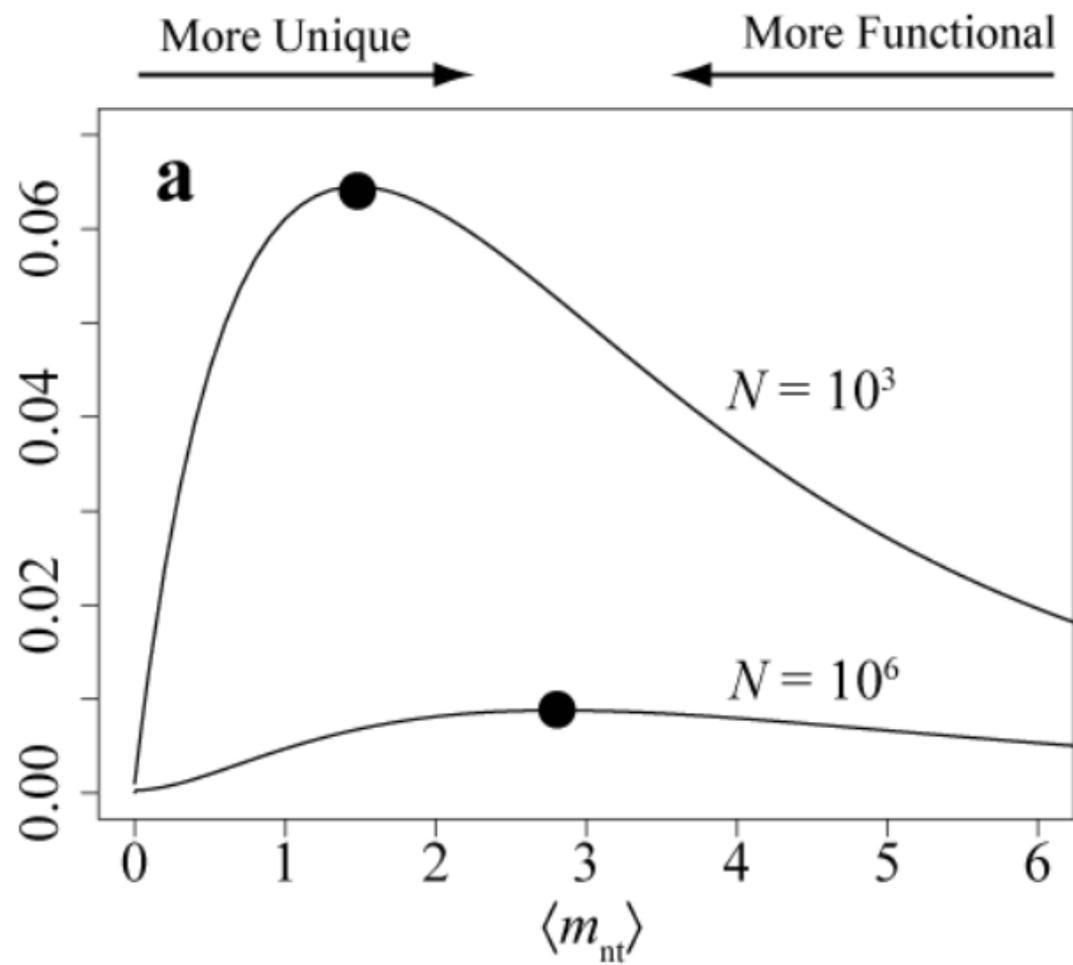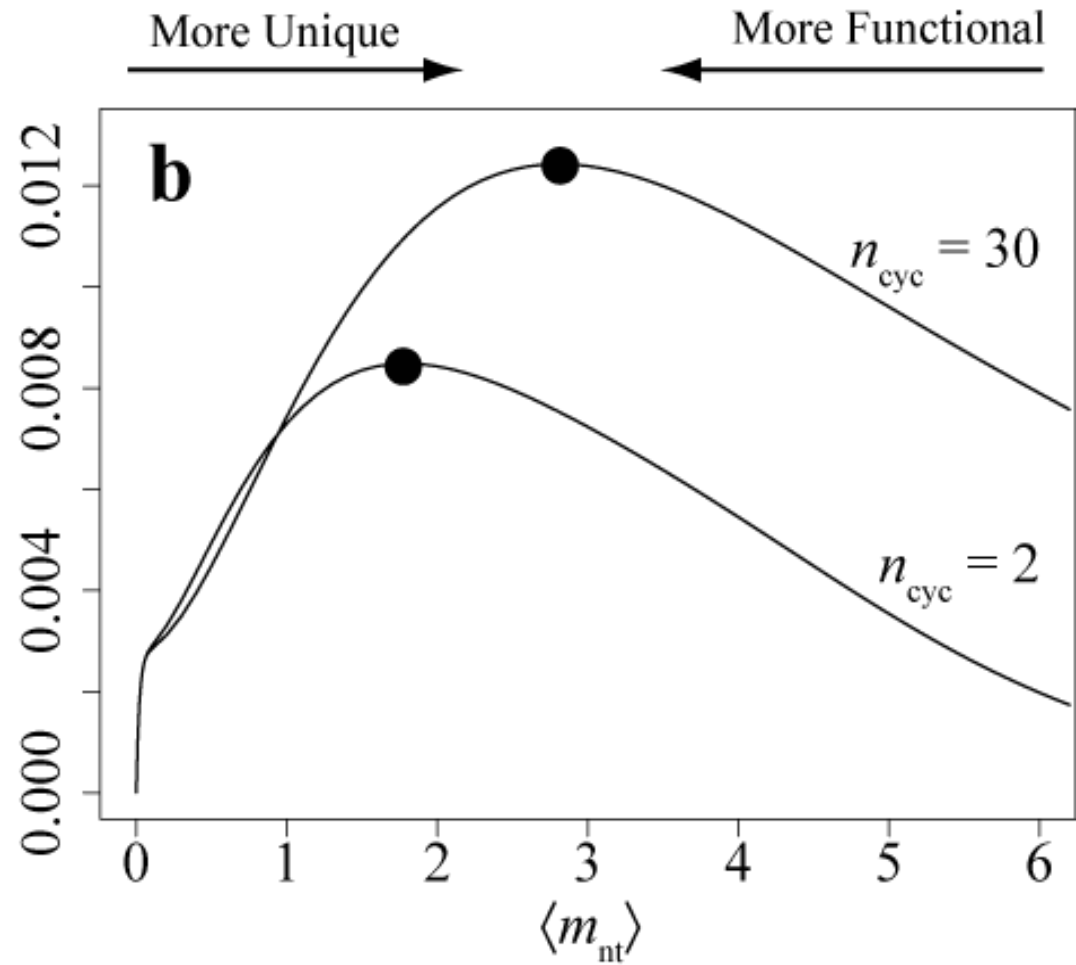

Table 1: scFv antibody mutational results and corresponding predictions for PCR and Poisson-distributed mutations.

| Library | # seq'd | $\langle m_{nt} \rangle$ | $\sigma^2_{m_{nt}}$ ($P(\sigma^2_{m_{nt}})$ if Poisson) | PCR $\sigma^2_{m_{nt}}$ [a] | Poisson $\sigma^2_{m_{nt}}$ |
|---|---|---|---|---|---|
| A | 46 | 15.8 ± 0.8 | 26.3 ($P < 0.005$) | 29.6 | 15.8 |
| B | 45 | 19.8 ± 0.9 | 36.1 ($P < 0.001$) | 41.4 | 19.8 |

[a] Assumed efficiency $\lambda = 0.6$ (18 DNA doublings).

**Table 2**: Mutational spectra [a] for libraries.

| Type | Library A (33,396 bp sequenced) | | Library B (32,670 bp sequenced) | |
| --- | --- | --- | --- | --- |
| | Number | Fraction | Number | Fraction |
| A→T, T→A | 172 | 0.24 | 106 | 0.12 |
| A→C, T→G | 7 | 0.01 | 7 | 0.01 |
| A→G, T→C | 336 | 0.46 | 202 | 0.23 |
| G→A, C→T | 188 | 0.26 | 529 | 0.60 |
| G→C, C→G | 11 | 0.02 | 28 | 0.03 |
| G→T, C→A | 11 | 0.02 | 17 | 0.02 |
| Total mutations | 725 | | 889 | |
| Nonsynonymous | 501 | 0.69 | 634 | 0.71 |
| Termination | 19 | 0.03 | 44 | 0.05 |

[a] In each gene, 726 nucleotides were sequenced. Sequences containing frameshift events were discarded, but occurred at a very low level (<5%).

**Table 3**: Comparison of retention of wildtype digoxigenin binding for scFv antibody libraries with analytical predictions.

| $\langle m_{nt} \rangle$ | $N$ | Observed functional | Observed % funct. | Predicted % funct. [a] (Poisson) | Predicted % funct. [a] (Eq. 3) | Predicted $U_f$ |
|---|---|---|---|---|---|---|
| 1.7 | 3×10⁵ | 1.4×10⁵ | 40.0 | 36.1 | 35.6 | 2,687 |
| 3.8 | 1×10⁶ | 6.7×10⁴ | 6.7 | 10.2 | 12.8 | 9,639 |
| 15.8 [b] | - | - | 0.12 | 0.0076 | 0.094 | - |
| 19.8 [b] | - | - | 0.041 | 0.00069 | 0.028 | - |
| 22.5 | 6×10⁶ | 1×10⁴ | 0.17 | 0.00014 | 0.15 | 1,603 |

[a] Assumed scFv $\nu = 0.2$ (see text), efficiency $\lambda = 0.6$ for all but highest-$\langle m_{nt} \rangle$ library, for which we estimate efficiency $\lambda = 0.3$.

[b] Only fractions functional were recorded for these libraries.

# Supporting Information

To accompany Drummond, Iverson, Georgiou and Arnold (2005), "Why high-error-rate random mutagenesis libraries are enriched in functional and improved proteins."

To test our analytical results, we carried out simulations of error-prone PCR. Because we wished to accurately model the effect of mutations on proteins, yet do so in a tractable way, we used lattice proteins for our *in silico* work. These simplified models of proteins share relevant properties with real proteins (notably thermostability and mutational tolerance[1]), but can be folded and assayed in a fraction of a second.

## MATERIALS AND METHODS

We implemented a published 5×5 two-dimensional square lattice model[2,3] in which chains of $L$=25 residues fold into a maximally compact structure representing one of 1081 possible self-avoiding compact walks not related by symmetry. Residues are one of 20 amino acids, contact energies between nonbonded neighboring residues are computed using published values (Ref. 4, Table 3), and conformational energy is the sum of all contact energies for that conformation. Each simulation run begins with an arbitrarily chosen target conformation and a minimum stability (maximum free energy –5.0 $kT$). Proteins are defined as functional if they fold to this conformation with free energy at or below this value.

Our analytical work describes the effects of mutation on genes of several hundred base pairs, the biologically relevant regime, but not on the 75bp genes encoding these lattice proteins due to the breakdown of the Poisson assumption. Thus we extended the protein model in a simple way: genes are 750 base pairs long and encode ten independently folding 25-residue "domains," initially identical in the wildtype, which must each fold to a target structure with the required free energy in order for the overall protein to retain fold.

Error-prone PCR was simulated as follows. Beginning with a set of 2000 identical template genes in the mix, sequences are duplicated with a probability equal to the PCR efficiency $\lambda$ and a per-site mutation rate $x = \frac{\langle m_{nt} \rangle (1+\lambda)}{n\lambda}$. This process is repeated for $n$ cycles. A sample of $N$ = 20,000 sequences is then taken of the resulting mix, translated according to the universal genetic code, and assayed for function according to the folding assay described above. The mutation rate was determined by sequencing these $N$ sequences; excellent agreement was found between the predicted rate $\langle m_{nt} \rangle$ and the actual rate, as well as with the standard error and that expected (see main text, Discussion; data not shown). The probability of truncation, $p_{tr}$, was set to 0.045; in this simulation, frameshifts do not occur, though stop codons do arise at a low frequency. The fraction of nonsynonymous mutations $p_{ns}$ was also determined from these sequences, and generally was in the range 0.7 to 0.8. The observed average value for each gene was used when evaluating Equation 3.

The number of unique genes, unique proteins, functional proteins, and unique and functional proteins was tabulated for each sample.

Because PCR is an exponential-growth process, simulation is notoriously difficult. We implemented an efficient simulation allowing us to obtain libraries at high mutation rates of >$10^6$ sequences on a modest desktop PC with a 2.8GHz Intel Pentium IV processor and 500MB of RAM. Performance is significantly better at low mutation rates due to the nature of the optimization (storing only mutational changes rather than entire sequences).

## RESULTS

**Fraction of functional proteins**

Using the protein model described in Materials and Methods, we found four genes encoding proteins with a wide range of $v$ values, from 0.13 to 0.8. We amplified these genes by simulated error-prone PCR per above. We also performed a mutagenesis run in which all mutations are introduced at once, the conditions under which a Poisson distribution of mutations should arise corresponding to the assumption made originally by Shafikhani *et al.*[5] discussed in the main text. Figure S1 shows the results of these simulations. The observed close agreement is typical and repeatable.

**Fraction of unique functional proteins**

Figure S2 shows the results of lattice-protein simulations compared to the simplified simulations described in the main text and with our theoretical results. The agreement is excellent and shows that essentially identical results can be obtained without a full simulation of the PCR process, as stated in the main text.

# FIGURES

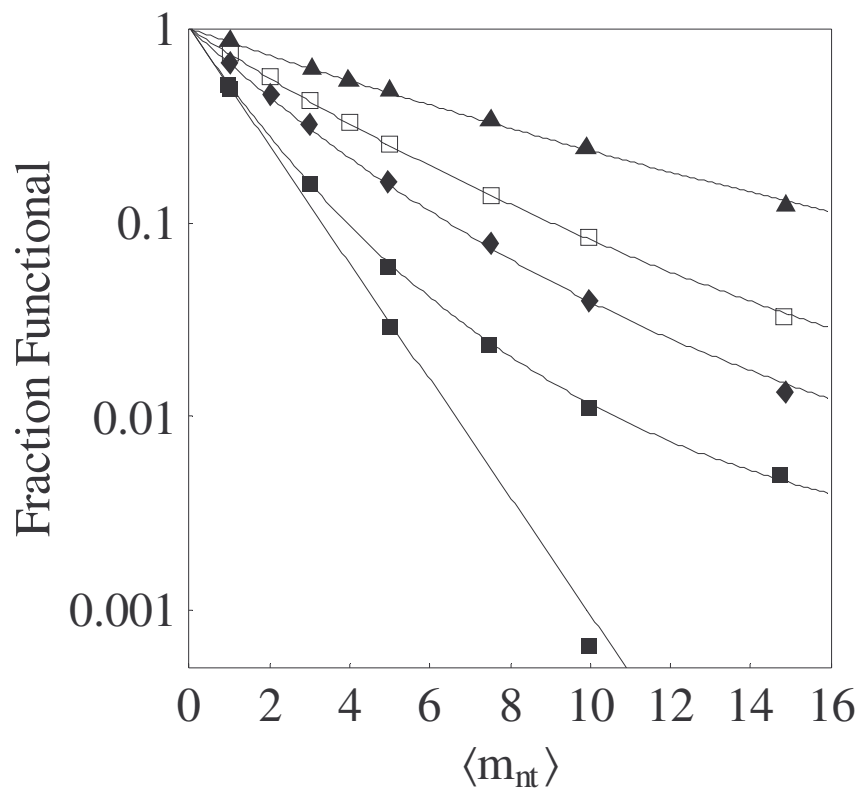

**Figure S1**: Comparison of Equation 3 to simulation results. Four proteins having domain structures with differing $v$ were assayed after error-prone PCR at $n = 16$ cycles, efficiency $\lambda = 0.5$. The lowest-$v$ structure was also subjected to single-round mutagenesis (Poisson-distributed mutations). The fraction of functional proteins is plotted (points) along with predictions using Equation 3 and, for the Poisson-distributed library, the equation $\Pr(f) = e^{-\langle m_{nt} \rangle (1-v) p_{ns}}$ (see main text).

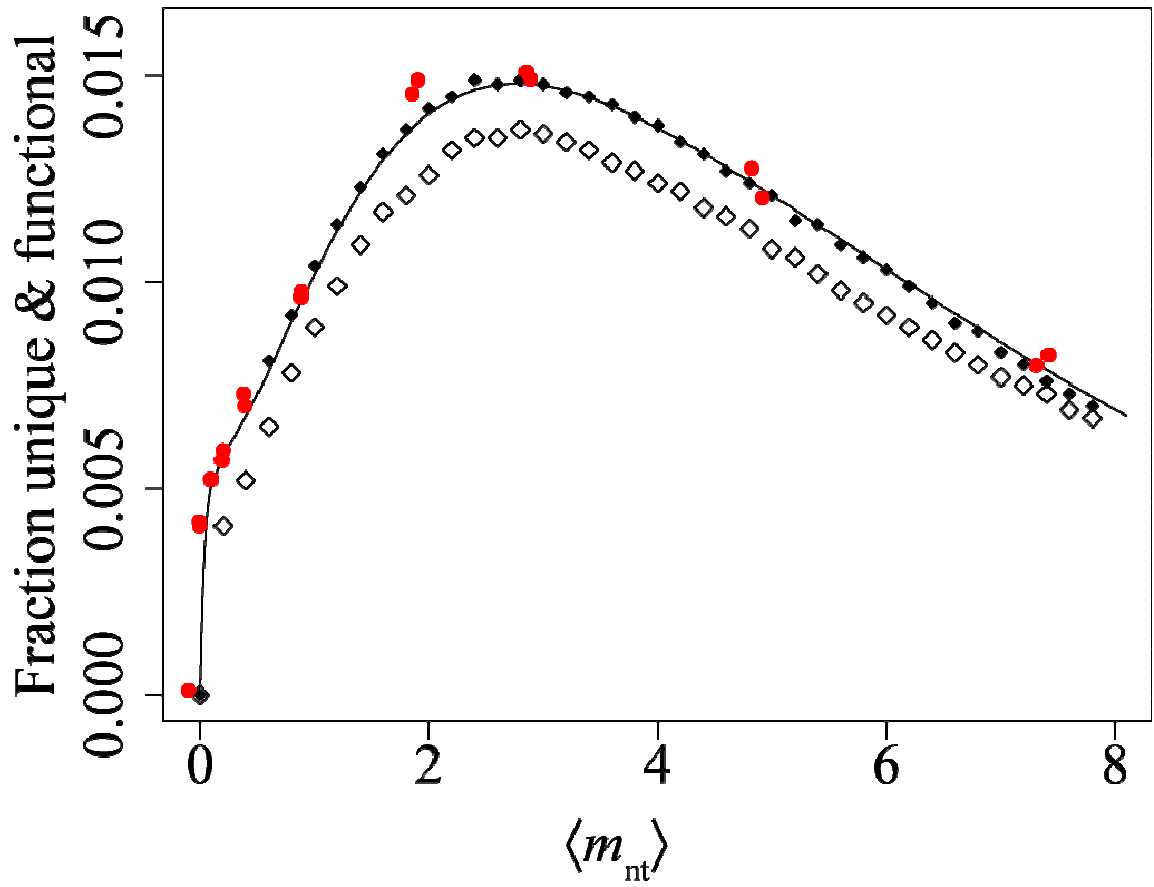

**Figure S2**: Comparison of simulation results to predictions for number of unique, functional proteins. Error-prone PCR conditions: $n$ = 14 cycles, efficiency $\lambda$ = 0.71, $\nu$ = 0.2, $p_{ns}$ = 0.76 and $p_{tr}$ = 0.07.